# DEMO: Simulation of Realistic Mobility Model and Implementation of 802.11p (DSRC) for Vehicular Networks (VANET)

Saurabh D. Patil
Student,
Terna Engineering College,
Navi Mumbai, India

D.V. Thombare
Associate Professor
Terna Engineering College,
Navi Mumbai, India

Vaishali D. Khairnar
Associate Professor
Terna Engineering College,
Navi Mumbai, India

## ABSTRACT
An ad hoc network of vehicles (VANET) consists of vehicles that exchange information via radio in order to improve road safety, traffic management and do better distribution of traffic load in time and space. Along with this it allows Internet access for passengers and users of vehicles. A significant characteristic while studying VANETs is the requirement of having a mobility model that gives aspects of real vehicular traffic. These scenarios play an important role in performance of VANETs. In our paper we have demonstration and description of generating realistic mobility model using various tools such as eWorld, OpenStreetMap, SUMO and TraNS. Generated mobility scenario is added to NS-2.34 (Network Simulator) for analysis of DSR and AODV routing protocol under 802.11p (DSRC/WAVE) and 802.11a. Results after analysis shows 802.11p is more suitable than 802.11a for VANET.

## Keywords
Reality mobility model, 802.11p, VANETs.

## 1. INTRODUCTION
In recent years, networks VANETs have had a great interest from researchers, whether from industry or academia. These networks were initially proposed for use in the military, today it has potential to be used for civil applications. MANET [2] network is a set of nodes interconnected by means of radio communication. These networks are distributed and are completely dynamic; in which each node must be able to configure itself without the need for any centralized management, or any infrastructure previously deployed. The similar featuring networks VANET which come as a subclass of MANET from outside. In which nodes are vehicles or terminals installed along the roads.

But in reality, networks VANET [1] are different by nature of their topology as it changes very frequently and very rapidly, nature of movement of vehicles, scope and scale of VANET networks. Their scope is related directly to human sight, the driver and passengers. VANET networks open a wide field of application in our daily lives. An application has to improve the comfort of drivers and passengers, and also to improve road safety and traffic management.

### 1.1 Defining Technology
An ad hoc network of vehicles or VANET consists of vehicles that exchange information via radio in order to improve road safety or to allow Internet access for passengers.

The integration of wireless communications in the new generation of intelligent transportation systems has become a priority. Specifically, it is one of objective of Intelligent Transportation Systems [11] (ITS, Intelligent Transportation System). Along with this we have many objectives such as

1. Reduce accidents (help the driver avoid accidents) by increasing the information available on the vehicle.
2. Better distribution of traffic load in time and space (help drivers spend less time on the roads, reduce traffic congestion).
3. Help the driver find what they need (a parking space, gas station, a gateway to the Internet, etc).

These characteristics and mobility support create new challenges such as time speed critical inter-vehicular communication to ensure road safety, the large number of vehicles, density up to several thousand in the same cell, the total coverage ( typically to ensure emergency service in case of accident). Addressing these challenges has been the main focus work in this area. Effectively routing traffic with a more intelligent use of resources is a good way to cover these problems. We can say that good design of the routing protocol is able to improve performance because the transmission and dissemination of information are made according to the protocol, the challenge we have already cited will be linked to the routing protocol (AODV and DSR) that will be used for VANET. Routing is an essential parameter in VANETs or this is the first parameters to be studied. So we need a comparison to be made between AODV and DSR under mobility Model.

The remainder of this paper is organized as follows. Section 2 describe the stepwise generation of realistic mobility model for VANET. Section 3 presents the setup of NS-2.34 for VANET using mobility model scenario generated and implementation of 802.11p (DSRC). Section 4 presents routing protocols in general and then specific routing protocol used for evaluation. Section 5 shows simulation scenarios and evaluation Results. The conclusion is provided in Section 6.

## 2. GENERATION OF MOBILITY MODEL
There are various numbers of open source and commercial tools and software available in market for generating traffic simulation model [3] for motion of multiple vehicles considering provided condition. In proposed paper we are using the OpenStreetMap (www.OpenStreetMap.org) for extracting road network in form of OSM file. Conversion of road network into XML format is done with help of eWorld, and then we use TraNS for executing the command of SUMO [9] for reading and simulating vehicular motion based on road network XML and traffic flow description.

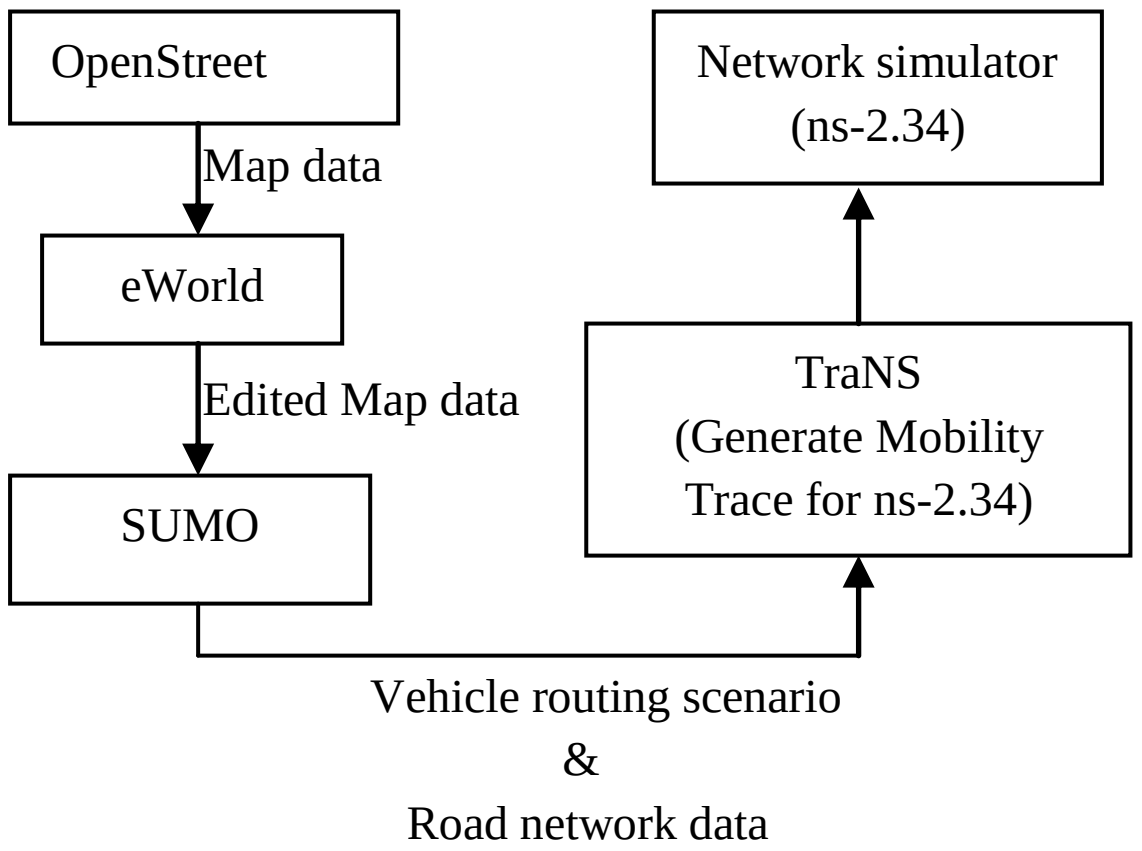

**Fig 1: Framework for generating .tcl file of realistic road map used in network simulator.**

In the proposed paper we have selected Palm Beach Road Vashi-India for mobility model. The stepwise generation of mobility model is given below:

1. Open www.OpenSteetMap.com for getting map data. Enter the location and select the appropriate road where you need to do the simulation. After selection of road map is done we need to select OpenStreetMap XML data and export the OSM file and save the same.
2. The OSM file will be imported in eWorld which is an open source project that provide the framework to convert map data from the OSM file to format that can be used for road traffic simulator including SUMO. The exported result of road network is given to SUMO in XML format.
3. The network and route XML files are generated by eWorld, this XML files are provide to TraNS. TraNS is a GUI based tools that help for integrating SUMO and ns2 for generating realistic scenario. Basically it converts SUMO mobility XML files to .tcl extension files use in ns2. Mobility traces are generated for ns2 from the net and rou XML files. The random and flow-based vehicle routes are generated using duarouter tool from SUMO.

In above step we need to configure TraNS by providing the bin folder path of SUMO and route generating tools path in preferences of TraNS.

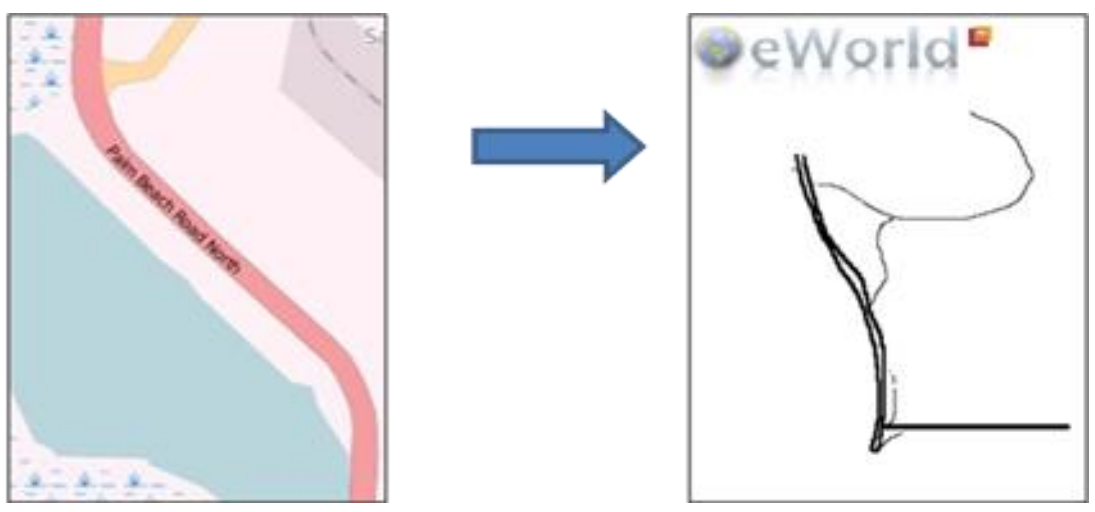

**Fig 2: OSM file of VASHI, Palm Beach Road in simulator Connector tool eWorld**.

Once we get the .tcl file we do the network simulation. For network simulation we are using ns2.34 version which support 802.11p.

## 3. IMPLEMENTATION OF 802.11P IN NS-2.34

As stated in [6], none of the current standards (802.11a) have completely adapted the new VANET technology. So IEEE had 802.11p [4] new protocol developed which can adapt properties of vehicular network and also support safety, quality and reliable data transmission in unstable network. The IEEE 802.11p, made formal changes to physical (PHY) and medium access control (MAC) layers, which is a modification to IEEE 802.11a for applying in the unstable vehicular environment.

In the physical PHY layer, minor changes made in number of channels, bandwidth and coding scheme. On the other hand, the functionalities of each channel and the enhanced priority control for traffic delivery are specified in the MAC part of the IEEE 802.11p. [6], [7].

In following table we have summary of parameters of 802.11a and 802.11p (DSRC) which help for better understand in short the difference between both the protocol. The parameters names are self explanatory.

**Table 1. 802.11p and 802.11a parameters Comparison**

| **Parameter** | **IEEE 802.11p** | **IEEE 802.11a** |
|---|---|---|
| Rate (Mbps) | 3, 4.5, 6, 9, 12, 18, 24 and 27 | 6, 9, 12, 18, 24, 36, 48 and 54 |
| Modulation | BPSK, QPSK 16-QAM and 64-QAM | BPSK, QPSK 16-QAM and 64-QAM |
| Codification Rate | ½, 1/3 and ¾ | ½, 1/3 and ¾ |
| Sub-carries Number | 52 | 52 |
| OFDM Symbol Duration | 8µs | 4µs |
| Guard Interval | 1.6µs | 0.8µs |
| FFT Period | 6.4µs | 3.2µs |
| Preamble Duration | 32µs | 16µs |
| Sub-carriers Spacing | 0.15624 MHz | 0.3125 MHz |

Here we will directly jump to the implementation part of 802.11p. For implementation we are using [5] NS (Network Simulator) version 2.34. VANETS support, however, was only developed in the last two versions (2.33 and 2.34). Its implementation is done by applying the IEEE 802.11p physical and MAC layers features in the TCL simulation code, defined by two native modules: ***WirelessPhyExt*** and ***Mac80211-Ext.***

| | |
|---|---|
| *set val(netif) Phy/WirelessPhy*<br>*set val (mac) Mac/802_11* | *set val(netif) Phy/WirelessPhyExt*<br>*set val(mac) Mac/802_11Ext* |
| By default setting of 802.11a | Changed setting for 802.11p |

## 4. ROUNTING PROTOCOL FOR EVALUATION

VANET Routing is a major challenge facing the preservation of the traditional distributed routing node in the database how to adapt to dynamic changes in network topology. Multi-hop Ad Hoc network routing is done in collaboration by the common node, rather than by a dedicated routing device completion. Therefore, we must design a dedicated and efficient wireless multi-hop routing protocol. At the moment, generally accepted representation of the results has been DSDV, AODV [8], DSR, TORA etc. So far, Routing Protocol in VANET is the most concentrated part of the results and we will only consider AODV and DSR as we are more focusing on reactive routing protocol.

DSR [10] protocol is a typical on-demand routing protocol, is the first thought of using on-demand routing protocols. DSR is based on the concept of on-demand source routing Adaptive routing protocols. Storage necessary to retain the mobile node of the source node know the route buffer. When the new route is found, the buffer entry is updated. Its biggest feature is the use of source routing mechanism; each packet header contains the entire road from the information.

AODV implicitly at each intermediate node stores the result of routing requests and responses, and using an expanding-ring search discovered a way to limit the search range of the destination node. AODV supports multicasting, support for QoS, and AODV can use the IP address, to achieve the same Internet connection, but do not support one-way channel. For setting up DSR/AODV we have set following line in tcl script.

***set val (rp) AODV/DSR ;***

## 5. RESULTS AND ANALYSIS

The evaluation and performance of routing protocol based on 802.11a and 802.11p, simulation will run on open source network simulator ns-2. Scenario characteristics are described briefly and then evaluation results are attempted which are obtained from these scenario.

**Scenario characteristics:** We are considering r realistic Scenario containing 1218 nodes (transmitter and receiver) which were defined, with 6282.22 m x 8829.25 m topology at a 2000 seconds simulation. The nodes movement was done in opposite directions and in each simulation the speed of each node was variable. The simulation was executed and the movement of vehicle was traced in .tr file. The next step was to test the routing protocol. We are using UDP transmission between 20 vehicles on same and opposite side of the road.

**Performance parameters:** To do analysis of AODV and DSR routing protocol, five main and very important performance parameters are considered:

1. Number of packets sent
2. Number of packets received
3. Packet delivery ratio
4. Point average delay
5. Receiving the first packet of the time

Packet delivery ratio: In order to evaluate the performance of AODV and DSR protocol by varying them between 802.11a and 802.11p. Fig.5.2 shows that Packet Delivery Ratio of AODV and DSR in 802.11a and 802.11p parameters is more as compare to 802.11p.

End to end delay: Average end to end delay is calculated and analyzed at Fig.5.3 and 5.4 by varying routing protocol under parameters of 802.11a and 802.11p. It is clear from results that delay in 802.11a is more compared to 802.11p and that is advantage of changing or adopting new protocol.

**Table 2. PDR results under 802.11p**

| | AODV | DSR |
|---|---|---|
| Number of packets sent | 261504 | 261504 |
| Number of packets received | 7200 | 9336 |
| Packet delivery ratio | 2.7533% | 3.57612% |
| Point average delay | 0.126377s | 0.107943s |
| Receiving the first packet of the time | 63.030412s | 35.146380s |

**Table 3. PDR results under 802.11a**

| | AODV | DSR |
|---|---|---|
| Number of packets sent | 261504 | 261504 |
| Number of packets received | 10435 | 24791 |
| Packet delivery ratio | 3.99038% | 9.48016% |
| Point average delay | 0.957310s | 1.365322s |
| Receiving the first packet of the time | 36.929281s | 19.181824s |

Table 2 represents the values of evaluation parameters under 802.11p where we have set routing protocol to AODV and DSR. The PDR is great with DSR protocol and even delay is less with DSR. But the execution time for formation of DSR trace is more as compare to AODV. Table 3 states about same parameters but under 802.11a even we see that PDR is great for both AODV and DSR as compared to 802.11p, important factor is End to End delay, where 802.11p is far better than 802.11a.

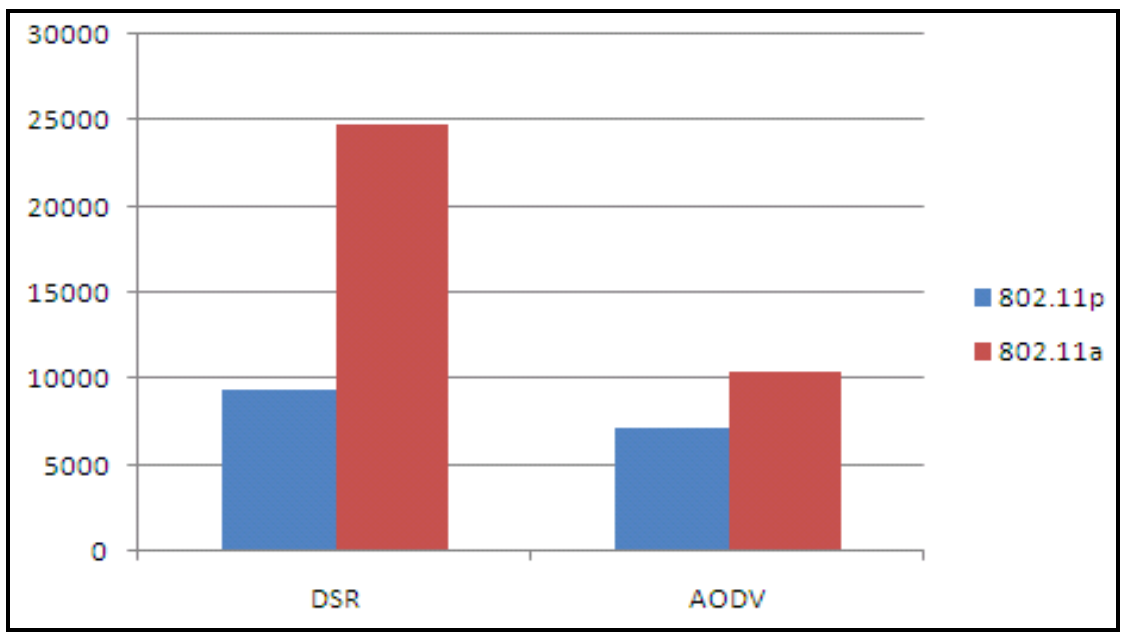

**Fig 2. Packet received on sending 261504 packets for DSR and AODV**

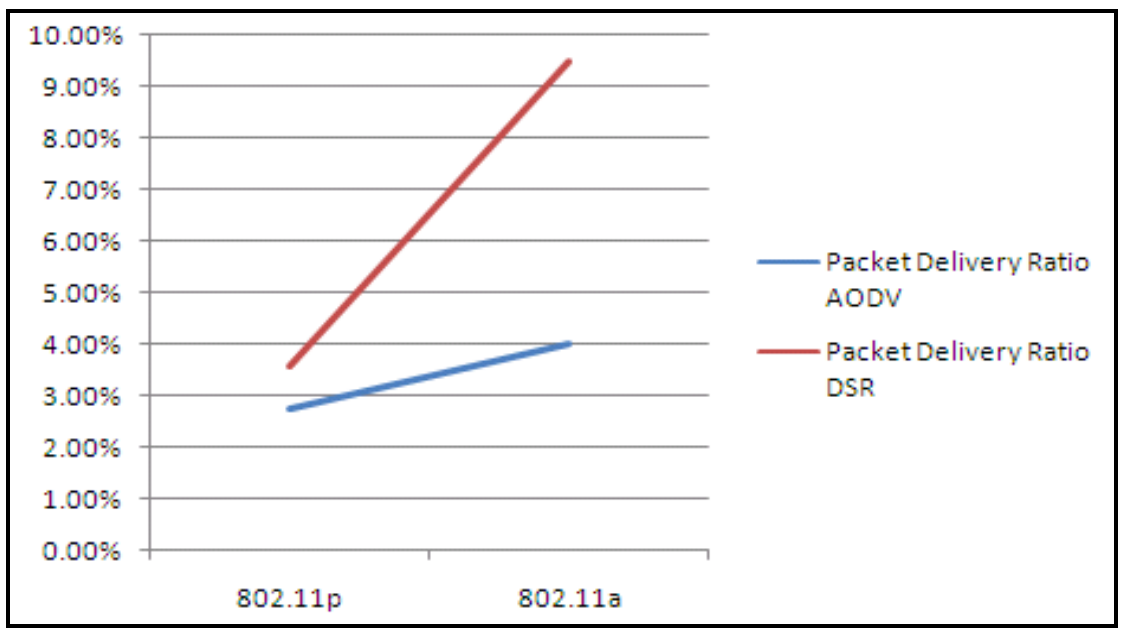

**Fig 3. PDR for AODV and DSR under 802.11p and 802.11a**

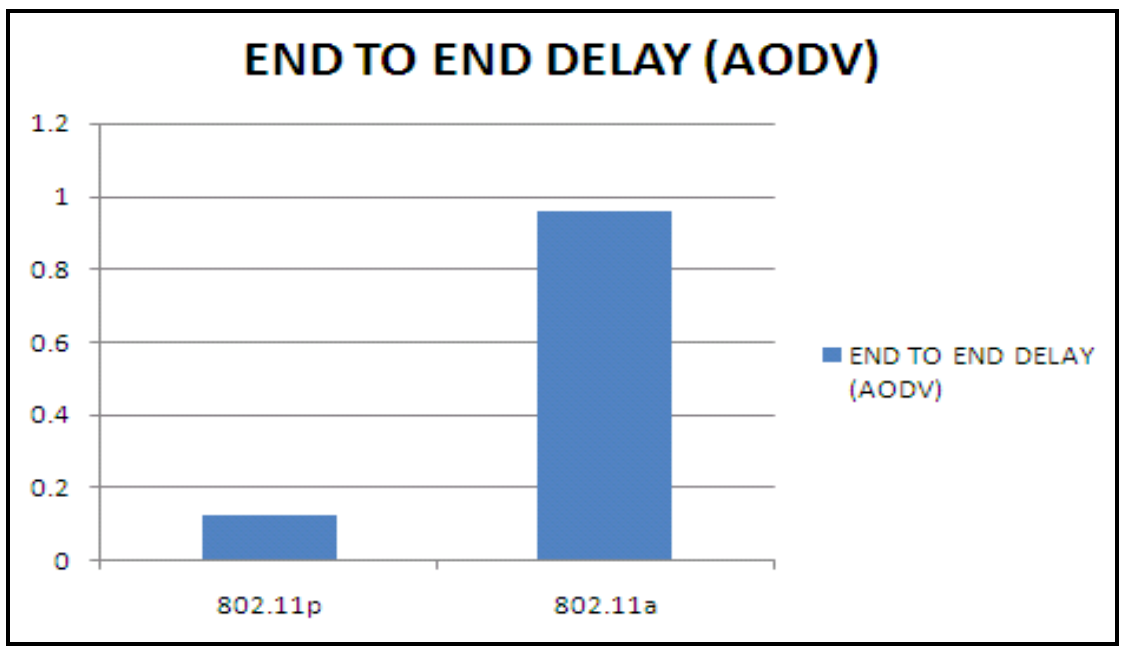

**Fig 4. End to End Delay for AODV under 802.11p and 802.11a**

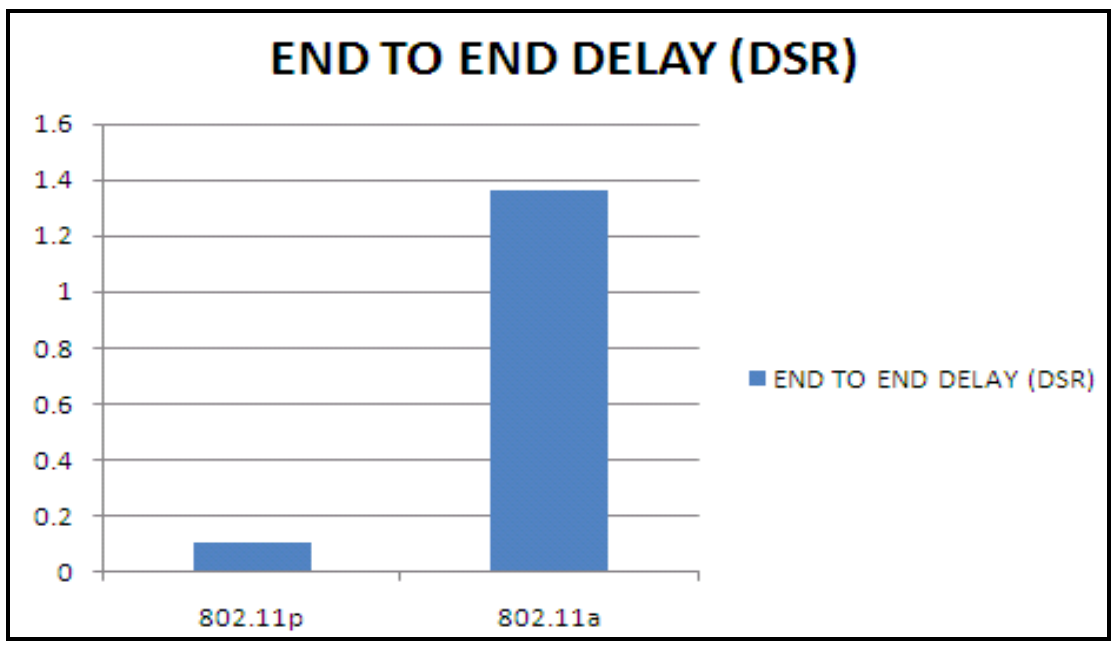

**Fig 5. End to End Delay for DSR under 802.11p and 802.11a**

Figure above shows the results of total packet send end to end delay and packet delivery ratio. We observed that the delay is more in 802.11a protocol as we have shown in Fig 5.

## 6. CONCLUSION

Researchers, government institute and mainly manufactures are looking forward for better and safe roads for driving. So this area of vehicular network has become more inserted topic and we have more development under VANET. Even IEEE has assigned dedicated IEEE 802.11p standard, which defines [12] the VANETS physical layer and MAC layer characteristics, real testbeds are still limited. This paper described how to generate the realistic mobility model and test computational experiment based simulations, using NS-2.34, whose main aim is to obtain the performance of two major routing protocols AODV and DSR by having 802.11p PHY and MAC parameters in VANETS. The results shows packets delay and packet delivery ratio of AODV and DSR under 802.11p and 802.11a. Our aim was to show the dedicated 802.11p have good results over 802.11a.